# FRAMR-EMR: Framework for Prognostic Predictive Model Development Using Electronic Medical Record Data with a Case Study in Osteoarthritis Risk

Jason E. Black, BMSc
Master's Student
Graduate Program in Epidemiology & Biostatistics
Western University
1151 Richmond Street
London, Ontario
Canada N6A 5C1
(jblack85@uwo.ca)

Amanda L. Terry, PhD
Assistant Professor
Department of Family Medicine, Department of Epidemiology & Biostatistics
Schulich Interfaculty Program in Public Health
Western University
1151 Richmond Street
London, Ontario
Canada N6A 3K7
(aterry4@uwo.ca)

Daniel J. Lizotte, PhD
Assistant Professor
Department of Computer Science, Department of Epidemiology & Biostatistics
Schulich Interfaculty Program in Public Health, Department of Statistical and Actuarial Sciences
Western University
London, Ontario
Canada N6A 3K7
(dlizotte@uwo.ca)

Corresponding Author:

Jason Black, BMSc
Graduate Program in Epidemiology & Biostatistics
Western University
1151 Richmond Street
London, Ontario
Canada N6A 5C1
(jblack85@uwo.ca)

**Abstract**

**Background:** Prognostic predictive models are used in the delivery of primary care to estimate a patient's risk of future disease development. Electronic medical record (EMR) data can be used for the construction of these models. Objectives: To provide a framework for those seeking to develop prognostic predictive models using EMR data, and to illustrate these steps using osteoarthritis risk estimation as an example.

**Methods (FRAMR-EMR):** The FRAmework for Modelling Risk from EMR data (FRAMR-EMR) was created, which outlines step-by-step guidance for the construction of a prognostic predictive model using EMR data. Our framework is composed of four main steps: Scoping and Background Search; Data Preparation; Cohort Construction; and Model Building and Evaluation. Throughout these steps, several potential pitfalls specific to using EMR data for predictive purposes are described and methods for addressing them are suggested.

**Results (Case Study):** We used the DELPHI (DELiver Primary Healthcare Information) database to develop a prognostic predictive model for estimation of osteoarthritis risk. We constructed a retrospective cohort of 28,447 eligible primary care patients. Patients were included in the cohort if they had an encounter with their primary care practitioner between 1 January 2008 and 31 December 2009; additionally, patients were required to have an encounter following the follow-up window to ensure patients were not lost to follow-up. Patients were excluded if they had a diagnosis of osteoarthritis prior to baseline. Construction of a prognostic predictive model following FRAMR-EMR yielded a predictive model capable of estimating 5-year risk of osteoarthritis diagnosis. Logistic regression was used to predict osteoarthritis based on age, sex,

BMI, previous leg injury, and osteoporosis. Internal validation of the model's performance demonstrated good discrimination and moderate calibration.

**Conclusions:** This study provides guidance to those interested in developing prognostic predictive models based on EMR data. The production of high quality prognostic predictive models allows for practitioner communication of accurately estimated risks of developing future disease among primary care patients. This allows practitioners and patients to create unique plans for addressing patient personal risk factors.

**Introduction**

Primary care is an ideal setting for the implementation of interventions to evaluate and reduce the risk of developing chronic disease [1,2]. Furthermore, the current level of electronic medical records (EMR) use in primary care [3–6] has sparked interest in using EMR-derived databases for research purposes [7], including risk modelling. Hence, the primary care setting presents a unique opportunity for developing risk models that are tailored to the very population who can make effective use of them.

Predictive models can be broken down into those that estimate the risk of disease presence (diagnostic) and those that estimate the risk of future disease development (prognostic) [8,9]. Prognostic predictive models estimate a patient's risk of future disease development based on that patient's predictors of disease [8,9]. Predictors may include patient demographics (such as age and sex), family history, lifestyle factors (such as smoking status or physical activity level), laboratory results, prior medical conditions, or genetic markers [10]. The estimates produced by a predictive model can be used in the patient care decision-making process of the primary care practitioner by informing patients and practitioners of future risk of disease [11,12]. Predictive models have been used to predict the development of many chronic diseases in patients [13–15], including osteoarthritis [16]. They have been effective in improving patients' risk perception and knowledge of risk [17] and in modifying primary care practitioner behaviour, including changes in practitioner prescribing practices [18].

Primary care EMR databases have several advantages over other data sources for use in predictive risk modelling:

- EMR data are often more abundant and less expensive to collect [19–21]
- They often represent a broader sample of the population compared to cohort studies that depend on volunteer participation [20]
- They often have longitudinal data spanning months or years
- When EMR data are used to build risk estimation tools, these tools can be more easily utilized in existing EMR software because they use risk indicators that are already present in EMR data [20] – an attribute that is greatly desired by primary care practitioners [22]

However, EMR data come with a distinct set of challenges:

- EMR data are more prone to have missingness compared to data collected specifically for research purposes [20,23,24]
- EMR data often lack standardization in terms of what data are collected, how they are collected, and how they are stored [20,21]

The TRIPOD (Transparent Reporting of a multivariable prediction model for Individual Prognosis Or Diagnosis) Statement [25] provides a comprehensive description of the methods required for developing predictive models; however, as it is a reporting guideline, the TRIPOD statement was not designed to guide predictive model development. To complement the TRIPOD statement, we present FRAMR-EMR, the FRAmework for Modelling Risk from EMR data, which is intended to serve as a guideline for researchers who wish to develop prognostic predictive models, with a specific focus on the use of EMR data. We then present a case study demonstrating how we applied FRAMR-EMR to EMR data to develop a prognostic predictive model for osteoarthritis.

**Methods (FRAMR-EMR)**

We developed the following framework based on the methods suggested by Moons et al. [25], Steyerberg [26], Lee et al. [10], and Hendriksen et al. [8]. Our framework extends this research, as we have developed our framework specifically for use with EMR data and have highlighted several concerns that are unique to EMR data. We intend for this framework to be used by researchers with a background in statistics but with possibly limited experience in predictive modelling who are interested in predicting disease outcomes. FRAMR-EMR comprises four main steps: Scoping and Background Search, Data Preparation, Cohort Construction, and Model Building and Evaluation. See Appendix 1 for a flow diagram of the framework.

1. Scoping and Background Search:

When developing a prognostic predictive model, researchers typically have a target population and a disease or health outcome in mind, which define the scope of the project. Given this scope, the next important step is a review of the literature to identify relevant risk indicators. We use the term *risk indicator* to mean any patient attribute that helps estimate risk. Risk indicators can be broken into: *risk factors*, indicators that have strong evidence for a causal effect; and *risk markers,* indicators that are associated but with unproven causality [27]. Identifying risk indicators *a priori* based on scientific and clinical knowledge helps avoid the discovery of spurious associations [8]. Having identified the potentially relevant risk indicators for the disease or outcome of interest, a collection of risk factors and markers should be identified as candidates for inclusion in the final model using the principles of analytic epidemiology [28].

With these risk indicators in hand, the next important decision is the selection of a data source. Our focus is the use of EMR databases as a data source. Many jurisdictions are developing aggregated primary care databases [29–32] that are available to primary care researchers. Different databases invariably contain different information, and may have had cleaning steps applied to address issues such as missing or implausible data. Some more well-developed databases may have been validated using methods such as chart review to ensure correctness of the data. It is important to understand what cleaning steps have been applied, the provenance of the data, and the characteristics of the population contained in the data source. The more closely the distribution of patients in the data source matches the population to which the resulting risk model will eventually be applied, the better the predictive performance will be. In the case of using primary care EMR data, the risk model may not be directly applicable to the general population as the data used to derive the model are not representative of the general population. However, they are representative of the primary care population, a common setting for risk estimation; thus, risk models derived from primary care data can be confidently applied to primary care populations.

2. Data Preparation and Definition:

An EMR is a rich source of information regarding a patient's medical conditions, health concerns, symptoms, laboratory results, physiologic measures, and medication history. From this record, we can extract data about a patient's disease and risk indicator status. A disease or risk indicator might be extracted as a binary variable (e.g., disease present or not present) or as a continuous value (e.g., BMI of 24) depending on how the pertinent data are stored in the database and on modelling choices. We strongly recommend developing precise disease case and risk indicator definitions that describe exactly how the disease or indicator status is derived from the available data. For simpler definitions, a set of decision rules may suffice [33]. When constructing these case definitions, it is best to draw on both written documentation of the database (e.g. a data dictionary [34]) as well as consultation with EMR users. One common way of identifying diseases is to examine the diagnosis codes within the billing records and problem list of the EMR. Consultation with EMR users is an important step to ensure that the extraction process makes valid assumptions about how disease and risk factor information are coded and stored within the EMR in practice. Ideally, the definition should be validated against a gold standard (e.g. chart review) where possible; however, given the evolving nature of this field, it is often not feasible to have a set of validated risk indicator definitions. Using EMR user consultation to develop the definitions helps ensure their validity when chart review is not possible. We strongly recommend validation of the disease outcome under study to avoid systematic over- or under-estimation of risk.

As the process of understanding the data, consulting with EMR-users, and developing definitions proceeds, it is useful to consider the dimensions of data quality of electronic health records (EHR) described by Weiskopf and Weng [35]: completeness, correctness, concordance, plausibility, and currency. EHRs and EMRs are very similar, thus these concepts pertaining to EHRs can easily be applied to EMRs. These concepts are defined as follows: "(1) Completeness: Is a truth about a patient present in the EHR?" (2) "Correctness: Is an element that is present in the EHR true?" (3) "Concordance: is there agreement between elements in the EHR, or between the EHR and another data source?" (4) "Plausibility: does an element in the EHR make sense in light of other knowledge about what the element is measuring?" (5) "Currency: is an element in the EHR a relevant representation of the patient state at a given point in time?". Assessing each of these dimensions as the project proceeds will strengthen the validity of the end result, and we refer the reader to the work by Weiskopf and Weng for a more detailed account. However, we note several issues specifically pertinent to EMR data under the dimensions of Completeness, Correctness (with Concordance and Plausibility as subordinate issues), and Currency.

*Completeness:* The role of EMRs is to facilitate the collection and storage of data that support the delivery of patient care [24,36], rather than research. Hence, data that are relevant to risk modelling but unimportant from a clinical care standpoint may be missing. In practice, dealing with partially missing data is a common problem; hence we discuss some specific mitigating strategies here. Where no data can be found regarding a particular risk indicator, it may be possible to link with external data sources to extract this information. (For example, risk indicators missing from a primary care EMR may be present in a hospital EMR, and could be linked using a unique patient ID.) If risk indicator data are present for some patients but not others, several different approaches may be appropriate. In the case of missing diagnoses (e.g. some patients have a diagnosis of asthma while most do not), it may be reasonable to assume that a missing diagnosis indicates that no disease is present. The validity of this assumption is dependent upon the disease or condition, and on the associated diagnostic and coding practices of primary care practitioners who populate the database; this once again underscores the need for effective user consultation. In the case of missing measurement data, such as physiologic measures or laboratory results, imputation methods may be appropriate for filling in these missing values [37]. However, great care must be taken as very often measurements will be missing because the practitioner has no reason to believe a measurement result would be abnormal and hence no reason to order the test; in that case, a "missing as normal" assumption may be appropriate [38]. Besides considering the reasons why data are missing, one must also consider whether there are sufficient original data for the results of multiple imputation to be reliable. We propose a simple simulation-based approach to assess the impact of imputation. Start by considering only the complete cases. If these are many, proceed by artificially eliminating data in a random fashion (but possibly depending on other data items) from the risk factor of interest to varying degrees. Repeat this procedure to create multiple datasets, each with a greater degree of missingness. For each of these artificially created datasets, impute the missing data using multiple imputation, and assess the ability of the imputation method to correctly fill in the missing data. This process will aid in determining to what extent data must be complete to provide reliable information for predictive purposes.

*Correctness:* Determining the correctness of our data would require a "gold standard" (a data source containing the true values of all data within the EMR); however, usually no such "gold standard" exists [39]. An examination of data *concordance* (whether or not data agree over time and across data sources) and *plausibility* (whether the data are physiologically and reasonably possible) is a useful way to identify incorrect data. It is often the case that multiple data elements in an EMR database provide information about the same risk indicator, because it is common for the same information to be recorded by the primary care practitioner in multiple areas of the EMR [24]. Where these data elements referring to the same risk factor disagree, these disagreements must be resolved. Determining which (if any) of these values is correct requires careful consideration aided by clinical insight. For physiologic values with known plausible ranges, it may be possible to identify and eliminate data that do not fall within these ranges [35]. This is a key part of data cleaning and should be applied to all such variables, because values outside plausible ranges are not only incorrect, but may also be outliers that influence undue (and incorrect) influence on estimated risk models. In this case, it may be better to delete the values and use an imputation method to compensate.

*Currency:* EMRs typically contain data reaching back to when the systems were first installed, and some contain records that have been scanned or manually entered from previous paper records. We stress the importance of ensuring that data are reasonably current for use, as there may be cohort effects on risk that could be mitigated by using up-to-date data. Criteria for determining whether the data are sufficiently current will vary depending on the disease and risk factors.

3. Cohort Construction

Our goal is to accurately model risk of developing the outcome of interest over a pre-specified time period. As patient data in the EMR are often organized into "encounters" or "visits" that record time-stamped data produced during an encounter with a practitioner, we suggest using these data to construct a retrospective cohort. Each patient will have a unique start-date within a given timeframe, at which the risk indicator and outcome definitions developed in Step 2 of our framework should be applied to determine their values. Eliminate patients who already have the outcome of interest at this point as these patients are not susceptible to the disease. To determine whether the patient developed the outcome of interest during the specified time period, we again apply the outcome definition to all encounter data recorded in the interval. If we find that the disease is present anywhere in the interval, we record the event and also possibly the time of the event, identifying that individual as a case. Individuals for whom we do not detect the event become non-cases. When working with EMR data, it is important to note that there is no way to know whether or not patients have been censored; that is, we cannot know for certain the status of the patient without record of an encounter as the patient may not have sought help for the disease or may have sought medical attention elsewhere. We suggest requiring an individual to have at least one observed encounter *after* the period of interest to confirm them as a non-case.

4. Model Building and Evaluation:

Retrospective cohorts constructed from EMRs as we have described above can be analyzed using many of the same techniques as other cohorts; however, given their typically much larger sample size, we make the following specific recommendations for building and evaluating models:

- Partition the cohort into training, development, and validation sets [40]. The training set will be used when building the predictive models. The development set will be used to assess the performance of each potential model and select the optimal model. The validation set—sometimes called the "test set" in machine learning literature—will be used to evaluate the performance of the final model. There is concern surrounding how large the development and validation sets should be. One method of determining the minimum sufficient size considers the c-statistic. In keeping with the TRIPOD guidelines [25], the c-statistic is a measure that is reported for the purposes of model validation. The c-statistic measures discrimination, the ability of the model to assign greater risk to a patient who will develop disease compared to one who will not, given their risk factors. A c-statistic of 0.5 means the model is no better than randomly assigning greater risk; any value greater than 0.5 means the model is performing better than random. The minimum partition size for the development and test sets to detect a significant difference from a c-statistic of 0.5 can be determined given a significance level, power, and kappa value (the ratio of controls to cases) [41]. For example, to detect a difference of 0.05 from random at a 5% significance level, 80% power, and kappa value of 10 (corresponding to ten times the number of controls as cases), 274 cases and 2737 controls would be required. Hence, EMR databases are typically large enough that methods such as cross validation and bootstrapping are not necessary.

- Modelling choices will first be informed by the form of the outcome. Given a continuous outcome, linear regression is often used. Given a binary outcome, logistic regression is often used. Given time-to-event data, Cox regression is often used. These initial methods can be used to explore the relationships between the predictors and the outcome; more advanced methods, such as generalized additive models, may be used where appropriate. For example, non-linearity is better captured by using generalized additive models [42], which is possible without overfitting due to the large sample sizes of EMR databases that support such techniques; this method should be used when researchers suspect the relationship between risk indicator and outcome is non-linear on the log scale. Other non-parametric methods exist, such as decision trees or k-nearest neighbours, which may fit the data better or be more in line with researcher objectives (e.g. decision trees are easily interpreted, which may be desired by researchers). Use these alternative methods where appropriate.

- Assess the performance of models produced by examining the discrimination and calibration of models on the development set. As mentioned, discrimination can be assessed visually using an ROC curve or objectively using the c-statistic, which is equivalent to the area under the ROC curve (AUC) [8]. Calibration is a measure of how well a model fits the given data. One method of assessing calibration plots the average estimated risk within each risk decile against the actual proportion of patients who develop the disease within the decile [4]. Ideally, the estimated risk should correspond to the observed rate of disease within the validation set. Another method of assessing calibration uses the Hosmer-Lemeshow test [43]; however, this test is over-sensitive when applied to large sample sizes and thus is not suitable for working with EMR data [10]. Based on these criteria, the best model can be selected.

- Once the best model is selected, rebuild the model using both the training and development sets to make use of all available data. As recommended by TRIPOD [25], evaluate the chosen model using the validation set and report these validation measures; this is referred to as internal validation. Internal validation of a model assesses the performance of a model on a held out portion of the data. External validation assesses the performance of a model on a similar but different group of patients that were not used when developing the model [44]; external validation requires the use of an external dataset. At a minimum, a model must be validated internally by partitioning the dataset or using cross-validation methods. However, external validation is the strongest evidence that a model is a good predictor of disease.

**Results (Case Study: An Osteoarthritis Risk Model)**

In the following, we present our case study of constructing a prognostic predictive model for the 5-year risk of an osteoarthritis diagnosis. As the most common joint disorder worldwide [45], osteoarthritis represents a growing concern for older adults [46]. Our prognostic predictive model enables the identification of patients at high risk of developing osteoarthritis, which in turn allows health and lifestyle modifications aimed at reducing the risk of disease development to be targeted to those with highest risk [47,48]. Ultimately, we see this model being developed into a purpose-built tool, which can be used routinely by primary care practitioners during patient encounters to: 1) deliver a quantitative assessment of osteoarthritis risk in patients where the patient and/or primary care practitioner is concerned about osteoarthritis risk; and 2) act as a risk screening tool to detect high risk patients who may have gone undetected otherwise.

1. Scoping and Background Search

We sought to develop a prognostic predictive model for the diagnosis of osteoarthritis in Canadian primary care. We begin by examining existing literature to identify established risk indicators for the development of osteoarthritis. These were: BMI (Body Mass Index) [49–55], previous leg injury [49,51–53], leg length inequality [56], older age [50–52,54,55], female sex [50–52,55], osteoporosis [50], family history [55], occupation [55], and physical workload [54].

We used the DELPHI (Deliver Primary Healthcare Information) database to develop our prognostic predictive model. The DELPHI database contains de-identified patient records from the EMRs of 14 primary care practices throughout Southwestern Ontario, which includes more than 60,000 patients [29,57]. Within this database are all structured patient records, including: patient encounters, patient demographics, billing codes, laboratory results, prescriptions, referrals, risk factor information, and medical procedures. The DELPHI database contains only the structured data within the EMR; it does not contain the free text narrative. Several steps were previously carried out from the outset of database construction to ensure data quality, including: EMR user support; implementation that did not disrupt the workflow of EMR users; and data entry training [58]. Ultimately, we found that DELPHI contained data describing five of the nine risk indicators that we identified from the literature: BMI, previous leg injury, older age, female sex, and osteoporosis. Data describing family history, occupation, leg length inequality, and physical workload were not available.

2. Data Preparation

The DELPHI network contributes data to a Canada-wide EMR database called CPCSSN (Canadian Primary Care Sentinel Surveillance Network). Researchers working with the CPCSSN database have constructed and validated a case definition for osteoarthritis [59]. Billing codes, problem list diagnoses, laboratory results, and medication lists were used in creating the case definition. We used this disease case definition in our study. We created risk factor definitions for each risk factor; information was combined from multiple data elements to derive the case definitions, as seen below. These case definitions were constructed through examination of the DELPHI data dictionary and through consultation with physician end-users.

Table 1: Risk factor definitions for DELPHI database.

| Risk Factor | Table Name | Value |
|---|---|---|
| Age | Patient Demographics | Numeric |
| Sex | Patient Demographics | Male or female |
| BMI | Patient Encounter | Numeric |
| Leg Injury | Billing<br>Health Condition<br>Encounter Diagnosis | ICD-9 Codes:<br>• 820-29: fracture of lower limb<br>• 843: sprain or strain of hip and thigh<br>• 844: sprain or strain of knee and leg<br>• 928: crushing injury to lower limb |
| Osteoporosis | Billing<br>Health Condition<br>Encounter Diagnosis | ICD-9 Code:<br>• 733: Osteoporosis and other bone disorders |
| | Health Condition<br>Encounter Diagnosis<br>Risk Factor | "osteoporosis" |
| | Medications | Alendronic acid<br>Risedronic acid<br>Ibandronic acid |

The risk indicator "osteoporosis," as we have chosen to define it, includes any suspected bone disorder. This includes use of the ICD-9 (International Classification of Diseases – Ninth Revision) code 733 (which is used to note osteoporosis and other bone disorders) in the problem list, billing data, or encounter diagnosis fields; the term "osteoporosis"; or prescription of a medication commonly used for the treatment or prevention of osteoporosis. Specific diagnoses of osteoarthritis (ICD-9 code 733.0) were not available.

We then examined the DELPHI database in regards to the quality of its data. We considered data concordance; however, information was only found in one location for each risk factor. Not all data were plausible: some patients had a birth year of 0; we considered these values to be missing. We also considered BMI measurements greater than 100 kg/m$^2$ or less than 10 kg/m$^2$ to be missing, as these values were deemed implausible. Risk indicator data for leg length inequality, physical workload, occupation, and family history were completely missing, and we were not able to obtain them by linking with external databases. Information pertaining to age was missing in roughly 15% of patients while information pertaining to BMI was missing in roughly 28% of patients. We used multiple imputation to complete these data.

The DELPHI database contains all visit data for the participating practices up to 21 January 2016; hence, we consider the information to be current.

3. Cohort Construction

The construction of our cohort is illustrated in Figure 1. We included any patient who had a visit with their primary care practitioner between 1 January 2008 and 31 December 2009 and had not been previously diagnosed with osteoarthritis. Risk indicators were assessed on the date of this visit by examining records prior to and including this date. Patients were then "followed" in the database for 5 years. Those who were diagnosed with osteoarthritis within this time period and who had at least one recorded visit after the 5-year period were considered positive cases of osteoarthritis. The additional recorded visit was required to ensure that they were not lost to follow-up/censored. We also eliminated any patient who was diagnosed with osteoarthritis at their follow-up visit. This was done as we could not be sure that this diagnosis would not have been present prior to the end of follow-up had they visited their primary care practitioner. Our final dataset included 28447 patients.

Figure 1: Flowchart depicting patients included in retrospective cohort.

Having identified our cohort, we then developed and applied our risk indicator definitions. We began by determining a BMI value for each patient. Because many patients did not have a measurement recorded at intake, we used interpolation of adjacent values to estimate BMI at intake. Where there was only a measurement on one side of the intake visit (either before or after, not both) we used the closest value.

Table 2: Characteristics of our research sample

| Characteristics | Research Sample |
| --- | --- |
| Sample size | 28447 |
| Age, mean (SD), years | 42.7 (21.8) |
| Females, % | 55.2% |
| BMI, mean (SD), kg/m$^2$ | 28.1 (7.9) |
| Patients with leg injuries, % | 4.2% |
| Patients with osteoporosis, % | 2.1% |

SD, Standard Deviation; BMI, Body Mass Index

We then performed multiple imputation using our resulting cohort. In order to produce the best imputation possible, we included all identified risk indicators as well as systolic blood pressure, and number of chronic diseases; systolic blood pressure and number of chronic diseases were included to improve the imputation, but were not included in the predictive model as they are not risk indicators for osteoarthritis. We imputed 20 datasets using the MICE package in R [60].

4. Model Building and Evaluation

We began by randomly partitioning our cohort of patients into training, development, and validation sets. We then built several predictive models:

- Logistic regression
- Logistic regression with continuous predictors logarithmically transformed
- Logistic regression with the addition of quadratic transformations of the continuous predictors
- Generalized additive models
- Generalized additive models with continuous predictors logarithmically transformed

Each model was built from each of the 20 imputed training datasets separately, then combined using Rubin's rules [61]. We used both logistic regression (using the R package: *stats* [62]) and generalized additive models (using the R package: *mgcv* [63]). We also applied logarithmic transformations to the positive-valued continuous variables: age and BMI. To evaluate each combined model, we applied them to the development sets, resulting in 20 measures of performance, which were also combined, again using Rubin's rules.

After comparing the performance of the different models on the development set, we found that the simplest model – logistic regression – appeared to be the best. Having selected this model, we re-estimated the parameters using both the training and development set, which produced our final model. The performance of our final model was then assessed using the same performance measures on the reserved validation set.

The following model was produced:

*Logit = -5.29 + 0.04 (Age) + 0.14 (Sex) + 0.02 (BMI) + 0.36 (Leg Injury) + 0.60 (Osteoporosis)*

Discrimination: The area under the ROC curve showed the model had a good discriminative ability (AUC 0.74, 95% confidence interval: 0.71 to 0.76).

Calibration: Model calibration is illustrated in Figure 2. Calibration appears good for low-risk patients. The model tends to overestimate risk in the upper deciles, then underestimates risk at large values in the extreme upper deciles.

Figure 2: Calibration plot of logistic regression model's performance on test set.

**Discussion**

Prognostic predictive models serve several purposes in the clinical setting: screening of individuals to identify high risk patients; prediction of future morbidity; and assistance in primary care practitioners' clinical decision making [10]. However, the construction of a prognostic predictive model is no simple task; there is no single method of developing these models. Our framework, FRAMR-EMR, aims to provide guidance for researchers aspiring to construct such models using EMR data.

Construction of a prognostic predictive model following FRAMR-EMR yielded a predictive model capable of estimating 5-year risk of osteoarthritis diagnosis. Internal validation of the model's performance demonstrated good discrimination and moderate calibration. Compared to previous osteoarthritis risk estimation models [16,64,65], our model is more easily applied in clinical practice as it requires no additional measurements by the physician beyond what is already stored within the EMR; previous models require information such as occupational status or a Kellgren and Lawrence (K&L) score based on radiographic images. By exclusively using predictors that are already contained within the EMR, our tool will be able to operate in the background of the EMR, flagging patients during a patient encounter that are at high risk of developing osteoarthritis. By identifying these patients at high-risk of osteoarthritis, this simple tool will allow physicians to be aware of which patients should be closely monitored for disease development and for whom strategies to reduce risk, such as weight loss [66], should be prescribed. We foresee this model aiding in the decision making process of physicians to help delay or prevent the development of osteoarthritis in high-risk individuals.

Existing methods of predictive model development typically assume that the data to be used for predictive model development are in a readily usable form [8,10,11]; this is consistently not true for EMR data [23] as they are created for clinical, not research use [67]. Our work aims to address this by discussing ways in which data may be less than ideal and providing methods of handling and cleaning such data. Missing data, for example, is an issue of particular significance in EMRs; we have suggested methods for dealing with missing data. We also address how to construct a cohort for the development of predictive models, since the cohort must be defined and constructed from the EMR data alone. Similar work, such as that of Ferrão et al., has presented general guidelines for preprocessing patient record data to convert it into a useable form for research; in contrast, our work presents a comprehensive framework specifically for developing a prognostic predictive model using primary care EMR data. The FRAMR-EMR describes all the steps necessary to move from raw patient data to a functioning predictive model.

The use of EMR data for prognostic predictive modelling has many advantages over other data sources, including the size and amount of detail contained within these databases; however, there exist many limitations that must be considered as well. Our framework has been developed specifically for the use of EMR data alone; we did not explicitly consider other data sources except through linkage. EMR data often do not contain information regarding potentially highly predictive risk indicators, such as lifestyle factors, which are not regularly recorded in an EMR. For example, physical workload has been shown to be highly associated with development of osteoarthritis [54]; however, information regarding physical workload was not contained within the DELPHI database.

Conclusions

Our goal was to produce a useful framework for researchers and analysts interested in developing prognostic predictive models based on EMR data, and to illustrate its use on a specific, real-world predictive modelling problem. Our hope is that the increased production of high quality prognostic predictive models will enable communication of accurately estimated risks of developing future disease, and in turn will help to tailor interventions to individual patients' risk profiles and reduce the overall burden of disease.

List of Abbreviations:

EMR: electronic medical record

FRAMR-EMR: FRAmework for Modelling Risk from EMR data

DELPHI: DELiver Primary Healthcare Information

AUC: area under the ROC curve

BMI: body mass index

CPCSSN: Canadian Primary Care Sentinel Surveillance Network

ICD-9: International Classification of Diseases – Ninth Revision

**Declarations**

Ethics: Ethics approval was obtained from the Western University Research Ethics Board #107572.

Consent for publication: not applicable as no individual person's data were presented

Availability of Data and Materials: The DELPHI database is not publically accessible, in keeping with intent of the agreement made with primary health care practitioners contributing to the DELPHI database. As such, these data are currently available to researchers who are linked with the DELPHI team, following a data request procedure.

Competing Interests: the authors have no competing interests to declare.

Funding: Funding for this research was provided by NSERC. The funding source played no role in this research.

Authors' Contributions: Lead author was JB. All authors initiated the research idea. JB drafted the research idea. JB wrote the article that is being submitted. AT and DL revised the article. JB extracted and analyzed the DELPHI data. DL and AT supported the methodology development and contributed to the writing of the article. Each author has read and approved the final version of this article.

Acknowledgements: The DELPHI (Deliver Primary Healthcare Information) project was funded by the Canada Foundation for Innovation, the Primary Health Care Transition Fund, and the Enhancing Quality Management in Primary Care Initiative of the Ontario Ministry of Health and Long-Term Care. The views expressed here are those of the authors and do not necessarily reflect the views of the Ontario Ministry of Health and Long-Term Care.

and Over the Next 30 Years. 2011.

47. Felson DT, Zhang Y, Anthony JM, Naimark A, Anderson JJ. Weight loss reduces the risk for symptomatic knee osteoarthritis in women. The Framingham Study. Ann Intern Med [Internet]. 1992 Apr 1 [cited 2016 Jun 23];116(7):535–9. Available from: http://www.ncbi.nlm.nih.gov/pubmed/1543306

48. Felson DT. Weight and osteoarthritis. Am J Clin Nutr [Internet]. 1996 Mar [cited 2016 Jun 23];63(3 Suppl):430S–432S. Available from: http://www.ncbi.nlm.nih.gov/pubmed/8615335

49. Cooper C, Inskip H, Croft P, Campbell L, Smith G, McLaren M, et al. Individual risk factors for hip osteoarthritis: Obesity, hip injury and physical activity. Am J Epidemiol [Internet]. JOHNS HOPKINS UNIV SCHOOL HYGIENE PUB HEALTH, 111 MARKET PLACE, STE 840, BALTIMORE, MD 21202-6709 USA; 1998 Mar [cited 2015 Oct 9];147(6):516–22. Available from: http://apps.webofknowledge.com.proxy1.lib.uwo.ca/full_record.do?product=UA&search_mode=GeneralSearch&qid=7&SID=3DYVeRKRX5j7WUrx25B&page=1&doc=3

50. Lee KM, Chung CY, Sung KH, Lee SY, Won SH, Kim TG, et al. Risk Factors for Osteoarthritis and Contributing Factors to Current Arthritic Pain in South Korean Older Adults. Yonsei Med J [Internet]. YONSEI UNIV COLL MEDICINE, 50-1 YONSEI-RO, SEODAEMUN-GU, SEOUL 120-752, SOUTH KOREA; 2015 Jan [cited 2015 Oct 9];56(1):124. Available from: http://apps.webofknowledge.com.proxy1.lib.uwo.ca/full_record.do?product=UA&search_mode=GeneralSearch&qid=7&SID=3DYVeRKRX5j7WUrx25B&page=1&doc=1

51. Neogi T, Zhang Y. Osteoarthritis prevention. Curr Opin Rheumatol [Internet]. LIPPINCOTT WILLIAMS & WILKINS, 530 WALNUT ST, PHILADELPHIA, PA 19106-3621 USA; 2011 Mar [cited 2015 Oct 9];23(2):185–91. Available from: http://apps.webofknowledge.com.proxy1.lib.uwo.ca/full_record.do?product=UA&search_mode=GeneralSearch&qid=7&SID=3DYVeRKRX5j7WUrx25B&page=1&doc=2

52. Silverwood V, Blagojevic-Bucknall M, Jinks C, Jordan JL, Protheroe J, Jordan KP. Current evidence on risk factors for knee osteoarthritis in older adults: a systematic review and meta-analysis. Osteoarthritis Cartilage [Internet]. 2014 Nov 29 [cited 2015 Jan 17];23(4):507–15. Available from: http://www.ncbi.nlm.nih.gov/pubmed/25447976

53. Vignon E, Valat J-P, Rossignol M, Avouac B, Rozenberg S, Thoumie P, et al. Osteoarthritis of the knee and hip and activity: a systematic international review and synthesis (OASIS). Joint Bone Spine [Internet]. 2006 Jul [cited 2015 Oct 4];73(4):442–55. Available from: http://www.ncbi.nlm.nih.gov/pubmed/16777458

54. Vrezas I, Elsner G, Bolm-Audorff U, Abolmaali N, Seidler A. Case-control study of knee osteoarthritis and lifestyle factors considering their interaction with physical workload. Int Arch Occup Environ Health [Internet]. SPRINGER, 233 SPRING ST, NEW YORK, NY 10013 USA; 2010 Mar [cited 2015 Oct 9];83(3):291–300. Available from: http://apps.webofknowledge.com.proxy1.lib.uwo.ca/full_record.do?product=UA&search_mode=GeneralSearch&qid=7&SID=3DYVeRKRX5j7WUrx25B&page=1&doc=4

55. Leung GJ, Rainsford KD, Kean WF. Osteoarthritis of the hand I: aetiology and pathogenesis, risk factors, investigation and diagnosis. J Pharm Pharmacol [Internet]. WILEY-BLACKWELL, 111 RIVER ST, HOBOKEN 07030-5774, NJ USA; 2014 Mar [cited 2015 Oct 9];66(3):339–46. Available from:
30

Appendix 1

Figure 3: Flowchart of FRAMR-EMR.

List of Figures